\documentclass[%
 reprint,
 amsmath,amssymb,
 aip,
apl,
]{revtex4-1}

\usepackage{graphicx}
\usepackage{dcolumn}
\usepackage{bm}

\begin{document}

\preprint{APS/123-QED}

\title{Platinum thickness dependence of the inverse spin-Hall voltage from spin pumping in a hybrid YIG/Pt system }

\author{V. Castel}
 \email{v.m.castel@rug.nl}
\affiliation{ 
University of Groningen, Physics of nanodevices, Zernike Institute for Advanced Materials, Nijenborgh 4, 9747 AG Groningen, The Netherlands.
}%

\author{N. Vlietstra}
\affiliation{ 
University of Groningen, Physics of nanodevices, Zernike Institute for Advanced Materials, Nijenborgh 4, 9747 AG Groningen, The Netherlands.
}%

\author{J. Ben Youssef}%
\affiliation{ 
Universit\'e de Bretagne Occidentale, Laboratoire de Magn\'etisme de Bretagne CNRS, 6 Avenue Le Gorgeu, 29285 Brest, France.
}%

\author{B. J. van Wees}
\affiliation{ 
University of Groningen, Physics of nanodevices, Zernike Institute for Advanced Materials, Nijenborgh 4, 9747 AG Groningen, The Netherlands.
}%

\date{\today}

\begin{abstract}

We show the first experimental observation of the platinum (Pt) thickness dependence in a hybrid YIG/Pt system of the inverse spin-Hall effect from spin pumping, over a large frequency range and for different rf powers. From the measurement of the dc voltage ($\Delta\textrm{V}$) at the resonant condition and the resistance ($R$) of the Pt layer, a strong enhancement of the ratio $\Delta\textrm{V}/R$ has been observed, which is not in agreement with previous studies on the NiFe/Pt system. The origin of this behaviour is still unclear and cannot be explained by the spin transport model that we have used.

\begin{description}

\item[PACS numbers]
72.25.Ba, 72.25.Pn, 75.78.-n, 76.50.+g
\end{description}
\end{abstract}

\keywords{Yttrium Iron Garnet (YIG), liquid-phase-epitaxy, spin pumping, thickness dependence}
\maketitle

Recently in the field of spintronics, spin transfer torque and spin pumping phenomena in a hybrid ferrimagnetic insulator (Yttrium Iron Garnet: YIG)/normal metal (Platinum: Pt) system have been demonstrated\citep{Kajiwara2010nature}. Since this observation, the actuation, detection and control of the magnetization and spin currents in such systems have attracted much attention from both theoretical and experimental point of view.

Spin pumping in a ferromagnetic (NiFe)/normal metal system has been intensively studied as a function of the stoichiometric ratio between nickel and iron atoms\citep{SCNiFeratio}, the spin current detector material\citep{AndoIEEE2010}, and the ferromagnet dimensions\citep{1742-6596-266-1-012100,VariousSizeAndo2011}. Only a few research groups\citep{AzevedoPRB2011,NiFePtdep,Feng2012} have investigated experimentally and theoretically the dc voltage generation (induced by inverse spin-Hall effect) in a NiFe/Pt system as a function of the the thickness of the spin current detector and the ferromagnetic material, at a fixed microwave frequency. To date, no systematic studies of the spin/charge current (charge/spin) conversion in a YIG/Pt system have been presented as a function of the Pt thickness.

In this paper, we show the first experimental observation of the Pt thickness dependence in a hybrid system YIG/Pt of the inverse spin-Hall effect (ISHE) from spin pumping, actuated at the resonant condition by using a microstrip line in reflection over a large frequency range and for different rf powers.
  
\begin{figure}[h]
\includegraphics[width=8.5cm]{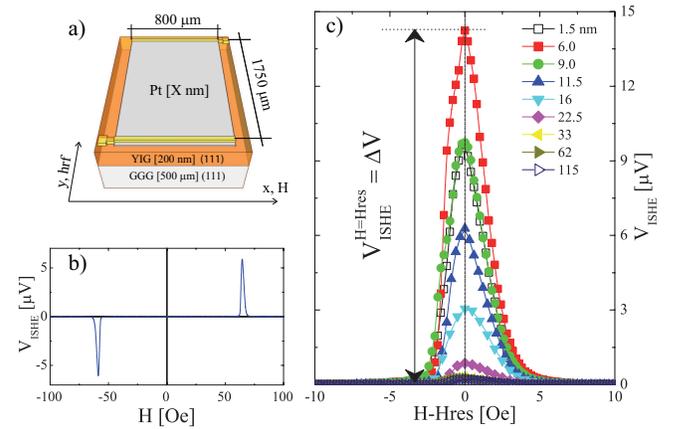}
\caption{\label{fig:Fig1} 
a) Schematics of the used sample for the inverse spin-Hall voltage detection. The magnetic field $H$ is applied in the plane of the sample along $\pm x$ directions and $H\perp h_{\texttt{rf}}$, where $h_{\textrm{rf}}$ is the microwave field applied by a microstrip. b) Measured voltage ($V_{\textrm{ISHE}}$) as a function of the magnetic field (positive and negative) at 1 GHz and 10 mW for a Pt thickness of 11.5 nm. c) Magnetic field dependence of the dc voltage for different thicknesses of Pt, measured at 1 GHz and 10 mW.
}
\end{figure}

The used insulating material consists of a single-crystal (111) Y$_3$Fe$_5$O$_{12}$ (YIG) film grown on a (111) Gd$_3$Ga$_5$O$_{12}$ (GGG) substrate by liquid-phase-epitaxy (LPE). We have prepared nine samples with different thicknesses of Pt (1.5, 6.0, 9.0, 11.5, 16, 22.5, 33, 62, and 115 nm) deposited by dc sputtering. Fig.\ref{fig:Fig1} a) shows a schematic of the samples. The Pt layer (800$\times$1750 $\mu$m) has been patterned by electron beam lithography (EBL). Ti/Au electrodes of 30 $\mu$m width and 100 nm thick have been grown on top of the Pt detector. The size of the YIG for each sample is 3000$\times$1500 $\mu$m with a thickness of 200 nm, which is very small for this kind of fabrication process (LPE).
Fig.\ref{fig:Fig1} b) shows the magnetic field dependence of the dc voltage in a YIG/Pt system with a Pt thickness of 11.5 nm for a rf power fixed at 10 mW at 1 GHz. At the resonant condition, $ H_{\textrm{res}} $, a spin current ($j_{s}$) is pumped into the Pt layer and converted in a dc voltage due to the ISHE. The reversal of the sign of $V_{\textrm{ISHE}}$, by reversing the magnetic field, shows that the signal is not produced by a possibly thermoelectrical effect induced by the ferromagnetic resonance (FMR) absorption. Fig.\ref{fig:Fig1} c) presents the magnetic field dependence of the dc voltage for different thicknesses of Pt measured at 1 GHz and 10 mW. As expected, the maximum value of the dc voltage, $\Delta\textrm{V}$, is reduced by increasing the Pt thickness. No significant changes of the shape of the dc voltage spectrum have been observed (also not at 3 and 6 GHz). 

For each thickness of Pt, we have analysed the dependence of the dc voltage induced by the ISHE as a function of the microwave power [0.25-70 mW], the frequency [0.1-7 GHz] and the in-plane static magnetic field, $H$, at room temperature.


\begin{figure}[h]
\includegraphics[width=8.5cm]{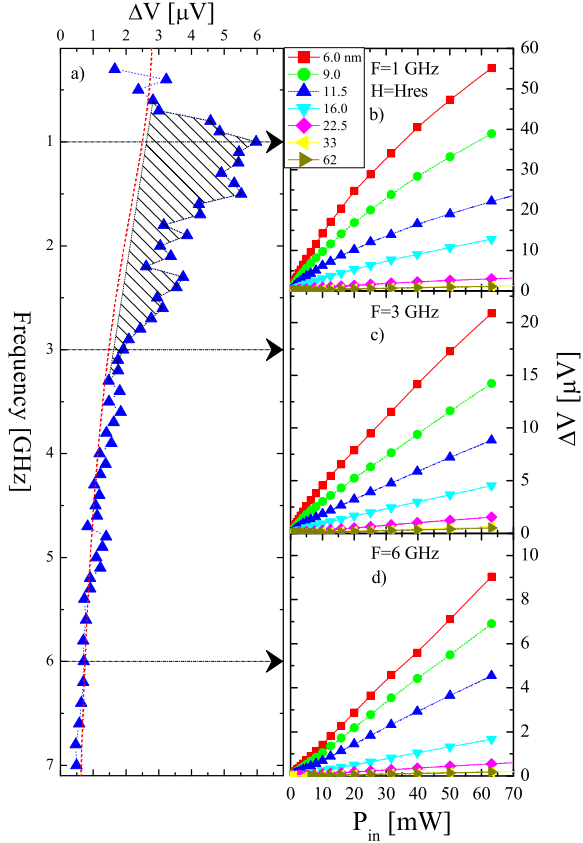}
\caption{\label{fig:Fig2} 
a) Frequency dependence of the dc voltage, $\Delta\textrm{V}$, at the resonant condition (see definition in Fig.\ref{fig:Fig1} c)) measured at 10 mW for a YIG/Pt system in which the Pt thickness is equal to 11.5 nm. The red dashed line corresponds to the theoretical expression extracted from Ref.\citep{Kajiwara2010nature}. b), c), and d) present the rf power dependence of $\Delta\textrm{V}$ at 1, 3, and 6 GHz, respectively, for different thicknesses of Pt. A strong non-linear behaviour has been observed at 1 GHz.
}
\end{figure}

The frequency dependence of $\Delta\textrm{V}$ measured at 10 mW for a Pt thickness of 11.5 nm is presented in Fig.\ref{fig:Fig2} a). The theoretical expression (red dashed line) extracted from Ref.\citep{Kajiwara2010nature} cannot reproduce the measured $\Delta\textrm{V}$ at low frequency. In Ref.\citep{Kajiwara2010nature}, the frequency dependence is defined by the magnetization precession angle, which is proportional to the ratio of the rf microwave field and the linewidth of the uniform mode. In other words, the spin current at the YIG/Pt interface is only defined by the damping parameter, $\alpha$, assuming that no spin waves are created\citep{Kurebayashi2011nmat,PhysRevLett.88.117601,PhysRevB.67.140404}. The enhancement of $\Delta\textrm{V}$ at low frequency has been observed for the set of samples with different thicknesses of Pt. As reported previously in Ref.\citep{HillebrandsSPmagnons, Kurebayashi2011nmat,YIGPt01}, this behaviour has been attributed to the presence of non-linear phenomena which are easily actuated for low rf power due to the very low damping of YIG. The dc voltage induced by spin pumping at the YIG/Pt interface is insensitive to the spin wave wavelength, which means that the measured $\Delta\textrm{V}$ is not only defined by the uniform mode (long wavelength) but also from secondary spin wave modes which present short wavelengths. 

For each sample, we have extracted $\Delta\textrm{V}$ as a function of frequency and rf power. A summary of the rf power dependence of $\Delta\textrm{V}$ at 1, 3, and 6 GHz is presented in Fig.\ref{fig:Fig2} b), c), and d), respectively. Each curve presents a different thickness of Pt, grown on top of the YIG sample. By decreasing the Pt thickness from 115 nm to 6 nm, we are able to detect dc voltages of 55, 21, and 9 $\mu$V at 1, 3, and 6 GHz respectively, for a rf power of 63 mW. The strong enhancement of $\Delta\textrm{V}$ for the thin layer of Pt (a factor of 70 between 6.0 and 115 nm of Pt) permits also to perform measurements at low rf power, lower than 250 $\mu$W (not shown).

In order to try to describe the Pt thickness dependence of $\Delta\textrm{V}$, observed in Fig.\ref{fig:Fig2} b), c), and d), we have derived the following expression. The general equation of the spin accumulation in the Pt layer is written as $\mu=a.e^{-z/\lambda}+b.e^{z/\lambda}$. From the boundary conditions, at z=0 nm and z=$t_{\textrm{Pt}}$, where $t_\textrm{Pt}$ is the Pt thickness of the spin current detector, one can write:

\begin{align}
   \begin{split}
       (a)~~~ & \left.  \frac{\textrm{d}\mu}{\textrm{d}z} \right|   _{z=t_\textrm{Pt}} =0    =-a\cdot e^{-t_\textrm{Pt}/\lambda}+b\cdot e^{t_\textrm{Pt}/\lambda}\\
       (b)~~~ &  b=a\cdot e^{-2t_\textrm{Pt}/\lambda}\\
       (c)~~~ & \mu(z=0)= \mu_{0}=a( 1+  e^{-2t_\textrm{Pt}/\lambda}) \\      
  \end{split} 
  		\label{Mu0}
\end{align}
where, $\lambda$ and $\mu_{0}$ correspond to the spin diffusion length of Pt and to the spin accumulation at the YIG/Pt interface, respectively. Therefore $\mu$ can be written as:
\begin{equation}
 \mu=\dfrac{\mu_{0}}{( 1+  e^{-2t_\textrm{Pt}/\lambda}) }
    	 [e^{-z/\lambda}+e^{-2t_\textrm{Pt}/\lambda}.e^{z/\lambda}]      
\end{equation}   		
The spin current, $j_{s}$, is written as:
\begin{align}
   \begin{split}
		& j_{s}=-\sigma\cdot\frac{\textrm{d}\mu}{\textrm{d}z}\\
	     j_{s} = \dfrac{\mu_{0}}{( 1+ e^{-2t_\textrm{Pt}/\lambda}) }& \cdot \frac{\sigma}{\lambda}
    	   \quad [e^{-z/\lambda}-e^{-2t_\textrm{Pt}/\lambda}.e^{z/\lambda}] 		
		\label{js}
  \end{split} 
\end{align}
where $\sigma$ corresponds to electrical conductivity of the Pt layer. From the definition of the spin current at the interface $j_{s}^{0}$ at z=0, the spin conductance of the Pt can be expressed as:
\begin{equation}
		g_\textrm{Pt}=\dfrac{j_{s}^{0}}{\mu_{0}}= \frac{\sigma}{\lambda}\cdot \frac{1-e^{-2t_\textrm{Pt}/\lambda}}{1+e^{-2t_\textrm{Pt}/\lambda}} 
	   		\label{gpt}
 \end{equation}
with $\mu_{0}\propto\dfrac{g_{\uparrow\downarrow}}{g_{\uparrow\downarrow}+g_\textrm{Pt}}$, where $g_{\uparrow\downarrow}$ corresponds to the mixing conductance of the YIG/Pt interface. The spin current at the YIG/Pt interface is then given by the equation: $j_{s}^{0} \propto \dfrac{g_{\uparrow\downarrow} \cdot g_\textrm{Pt}}{g_{\uparrow\downarrow} +g_\textrm{Pt}}$. From the equation of the spatially averaged spin current, given $\langle j_{s} \rangle=\dfrac{1}{t_\textrm{Pt}}\cdot \displaystyle{\int_{0}^{t_\textrm{Pt}}}j_{s}(z)~\textrm{d}z$, and the expression of the ISHE conversion of a spin current into a charge current, $\langle j_{c} \rangle=\left[  \frac{2e}{\hbar}\right]  \Theta_{\textrm{SH}}\cdot \langle j_{s} \rangle$, the dc voltage can be expressed as:
\begin{equation}
  V_{\textrm{ISHE}}= \dfrac{1}{t_\textrm{Pt}}\left[  \frac{2e}{\hbar}\right] L\Theta_{\textrm{SH}}\cdot\mu_{0}\cdot\frac{(1-e^{-t_\textrm{Pt}/\lambda})^2}{1+e^{-2t_\textrm{Pt}/\lambda}}
  \label{Vfull}
\end{equation}
where $\Theta_{SH}$ and $L$ denote the spin-Hall angle and the length (along $y$) of the Pt layer, respectively. Therefore, by combining equation (\ref{gpt}) and (\ref{Vfull}), the Pt thickness dependence can be written as:
\begin{equation}
V_{\textrm{ISHE}}\propto \dfrac{1}{t_\textrm{Pt}}\cdot \dfrac{g_{\uparrow\downarrow}}{g_{\uparrow\downarrow}+\frac{\sigma}{\lambda}\cdot \frac{1-e^{-2t_\textrm{Pt}/\lambda}}{1+e^{-2t_\textrm{Pt}/\lambda}} }\cdot \frac{(1-e^{-t_\textrm{Pt}/\lambda})^2}{1+e^{-2t_\textrm{Pt}/\lambda}}           
  		\label{V}
\end{equation}
Here we assume that the spin-Hall effect arises from extrinsic effects (scattering processes are dominant), and therefore $\Theta_{SH}$ should be independent of the Pt thickness\citep{RevPt}.

\begin{figure}[!htbp]
\includegraphics[width=8.2 cm]{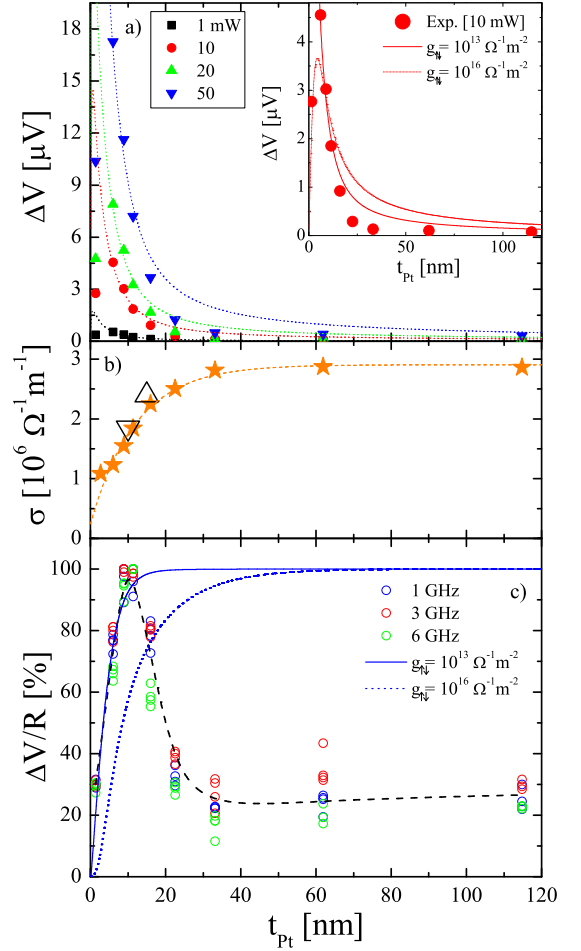}
\caption{\label{fig:Fig3} 
a) Pt thickness dependence of $\Delta\textrm{V}$ measured at 3 GHz. The rf power has been fixed at 1, 10, 20, and 50 mW. For each rf power, the dotted line shows the theoretical evolution of $\Delta\textrm{V}$ as a function of the Pt thickness, as given by Eq.(\ref{V}).The inset shows the dependence of $\Delta\textrm{V}$ as a function of the Pt thickness. The dots correspond to the experimental data at 3 GHz for a rf power of 10 mW. The solid line and the dotted line correspond to the theoretical dependencies from Eq.(\ref{V}) of $\Delta\textrm{V}$ when $g_{\uparrow\downarrow}$ is lower and higher than $g_\textrm{Pt}$, respectively. b) The dependence of the electrical conductivity, $\sigma$, of the normal metal is plotted as a function of the Pt thickness. The symbols $\bigtriangleup$ and $\bigtriangledown$ correspond to the magnitude of $\sigma$ extracted from Ref.\citep{PhysRevLett.104.046601} and Ref.\citep{2mag}, respectively. c) Pt thickness dependence of $\Delta\textrm{V}/R$ for different frequencies (1, 3, and 6 GHz) and rf powers (1, 10, 20, and 50 mW). The black dashed line corresponds to the average of the experimental data. The solid line and the dotted line correspond to the theoretical dependence of $\Delta\textrm{V}/R$ when $g_{\uparrow\downarrow}\ll g_\textrm{Pt}$ and $g_{\uparrow\downarrow}\gg g_\textrm{Pt}$, respectively.
}
\end{figure}

Fig.\ref{fig:Fig3} a) presents $\Delta\textrm{V}$ as a function of the Pt thickness, $t_\textrm{Pt}$, at 3 GHz, for different rf powers (1, 10, 20, and 50 mW). The general trend of $\Delta\textrm{V}$ is the same for the different rf powers and a maximum of $\Delta\textrm{V}$ between $t_\textrm{Pt}$=1.5 and $t_\textrm{Pt}$=6.0 nm has been observed (also at 1 and 6 GHz). Concerning the spin diffusion length, many values are reported, varying between 1.4 $\pm$0.4 (Ref.\citep{RevPt}) and 10 $\pm$2 nm\citep{PhysRevLett.104.046601}. In order to reproduce the strong dependence of $\Delta\textrm{V}$ for thinner layers of Pt, the best value obtained for the spin diffusion length of Pt has been found to be equal to $\lambda$=3.0 $\pm$0.5 nm, which is in good agreement with the value reported by A. Azevedo et al. \citep{AzevedoPRB2011}. For this value of $\lambda$, the spin conductance of the platinum, $g_\textrm{Pt}$, varies between 4.15 $10^{13}$ and 6.2 $10^{14}\Omega^{-1}\textrm{m}^{-2}$ at 1.5 and 11.5 nm, respectively. $g_\textrm{Pt}$ is constant for thicker layers of Pt, around 9.5 $10^{14}\Omega^{-1}\textrm{m}^{-2}$. The inset of Fig.\ref{fig:Fig3} a) presents the experimental dependence of $\Delta\textrm{V}$ at 3 GHz and 10 mW as a function of the Pt thickness including the theoretical dependencies from Eq.(\ref{V}) when $g_{\uparrow\downarrow}$ is lower (solid line) and higher (dotted line) than $g_\textrm{Pt}$. As observed in the inset of Fig.\ref{fig:Fig3} a), the mixing conductance should be lower than $g_\textrm{Pt}$ is order to reproduce partially the observed behaviour.

The electrical conductivity of the Pt layers (see Fig.\ref{fig:Fig3} b)) presents a strong Pt thickness dependence. $\sigma$ has been calculated from the measured resistance, $R$, and from the sample dimensions (the thickness of the Pt has been measured by Atomic Force Microscopy). $\sigma$ increases from 1.1 $10^{6}\Omega^{-1}\textrm{m}^{-1}$ (at 1.5 nm) to 2.9 $10^{6}\Omega^{-1}\textrm{m}^{-1}$ (at 33 nm). $\sigma$ presents a constant value between 33 to 115 nm ($\sim$2.9 $10^{6}\Omega^{-1}\textrm{m}^{-1}$). O. Mosendz et al.\citep{PhysRevLett.104.046601} and B. Jungfleisch et al.\citep{2mag} reported similar values of $\sigma$ for a Pt thickness of 15 and 10 nm, respectively, as shown in Fig.\ref{fig:Fig3} b).

The most interesting feature of this study is presented in Fig.\ref{fig:Fig3} c). This figure presents the dependence of $\Delta\textrm{V}/R$ as a function of the Pt thickness for different frequencies (1, 3, and 6 GHz) and rf powers (1, 10, 20, and 50 mW). The trend of these curves does not follow the behaviour observed in a NiFe/Pt system\citep{AzevedoPRB2011,NiFePtdep}. One would expect a constant value of $\Delta\textrm{V}/R$ for a Pt thickness higher than the spin diffusion length. Contrary to Ref.\citep{AzevedoPRB2011,NiFePtdep}, a maximum around 10 nm followed by a strong decrease until 20 nm of Pt has been observed. Values of $\Delta\textrm{V}/R$ are constant between 33 and 115 nm and the difference between the constant value of $\Delta\textrm{V}/R$ and the maximum is around 70$\%$. The solid line and the dotted line correspond to the theoretical dependence of $\Delta\textrm{V}/R$ when $g_{\uparrow\downarrow}$ is lower and higher than $g_\textrm{Pt}$, respectively. Two points should be made regarding these dependencies. First, when $g_{\uparrow\downarrow}\gg g_\textrm{Pt}$, the Pt thickness dependence of $\Delta\textrm{V}/R$ presents, for small thicknesses, a quadratic evolution and therefore cannot reach the experimental values. Second, when $g_{\uparrow\downarrow}\ll g_\textrm{Pt}$, the linear dependence of $\Delta\textrm{V}/R$ between 0 and 10 nm of Pt permits to reproduce partially the experimental behaviour. This shows that $g_{\uparrow\downarrow}$ should be lower than $g_\textrm{Pt}$, which is in agreement with results published by Y. Kajiwara et al.\citep{Kajiwara2010nature}. They have found that the mixing conductance at the YIG [1.3$\mu$m]/Pt [10 nm] interface is two orders of magnitude smaller than the computed value for a YIG/Ag system\citep{GerritMixCon}.

In conclusion, we have shown the first experimental observation of the Pt thickness dependence in a hybrid system YIG [200nm]/Pt [1.5-115nm] of the inverse spin-Hall effect (ISHE) from spin pumping, actuated at the resonant condition over a large frequency range and for different rf powers. A strong enhancement of the ratio $\Delta\textrm{V}/R$ has been observed for several frequencies and rf powers, which is different from previous studies on the NiFe/Pt system. The observed dependence of $\Delta\textrm{V}/R$ as a function of the Pt thickness in our system cannot be fully reproduced by our model. 

We would like to acknowledge B. Wolfs, M. de Roosz and J. G. Holstein for technical assistance. This work is part of the research program (Magnetic Insulator Spintronics) of the Foundation for Fundamental Research on Matter (FOM) and is supported by NanoNextNL, a micro and nanotechnology consortium of the Government of the Netherlands and 130 partners, by NanoLab NL and the Zernike Institute for Advanced Materials.


\begin{thebibliography}{17}%
\makeatletter
\providecommand \@ifxundefined [1]{%
 \@ifx{#1\undefined}
}%
\providecommand \@ifnum [1]{%
 \ifnum #1\expandafter \@firstoftwo
 \else \expandafter \@secondoftwo
 \fi
}%
\providecommand \@ifx [1]{%
 \ifx #1\expandafter \@firstoftwo
 \else \expandafter \@secondoftwo
 \fi
}%
\providecommand \natexlab [1]{#1}%
\providecommand \enquote  [1]{``#1''}%
\providecommand \bibnamefont  [1]{#1}%
\providecommand \bibfnamefont [1]{#1}%
\providecommand \citenamefont [1]{#1}%
\providecommand \href@noop [0]{\@secondoftwo}%
\providecommand \href [0]{\begingroup \@sanitize@url \@href}%
\providecommand \@href[1]{\@@startlink{#1}\@@href}%
\providecommand \@@href[1]{\endgroup#1\@@endlink}%
\providecommand \@sanitize@url [0]{\catcode `\\12\catcode `\$12\catcode
  `\&12\catcode `\#12\catcode `\^12\catcode `\_12\catcode `\%12\relax}%
\providecommand \@@startlink[1]{}%
\providecommand \@@endlink[0]{}%
\providecommand \url  [0]{\begingroup\@sanitize@url \@url }%
\providecommand \@url [1]{\endgroup\@href {#1}{\urlprefix }}%
\providecommand \urlprefix  [0]{URL }%
\providecommand \Eprint [0]{\href }%
\providecommand \doibase [0]{http://dx.doi.org/}%
\providecommand \selectlanguage [0]{\@gobble}%
\providecommand \bibinfo  [0]{\@secondoftwo}%
\providecommand \bibfield  [0]{\@secondoftwo}%
\providecommand \translation [1]{[#1]}%
\providecommand \BibitemOpen [0]{}%
\providecommand \bibitemStop [0]{}%
\providecommand \bibitemNoStop [0]{.\EOS\space}%
\providecommand \EOS [0]{\spacefactor3000\relax}%
\providecommand \BibitemShut  [1]{\csname bibitem#1\endcsname}%
\let\auto@bib@innerbib\@empty
\bibitem [{\citenamefont {Kajiwara}\ \emph {et~al.}(2010)\citenamefont
  {Kajiwara}, \citenamefont {Harii}, \citenamefont {Takahashi}, \citenamefont
  {Ohe}, \citenamefont {Uchida}, \citenamefont {Mizuguchi}, \citenamefont
  {Umezawa}, \citenamefont {Kawai}, \citenamefont {Ando}, \citenamefont
  {Takanashi}, \citenamefont {Maekawa},\ and\ \citenamefont
  {Saitoh}}]{Kajiwara2010nature}%
  \BibitemOpen
  \bibfield  {author} {\bibinfo {author} {\bibfnamefont {Y.}~\bibnamefont
  {Kajiwara}}, \bibinfo {author} {\bibfnamefont {K.}~\bibnamefont {Harii}},
  \bibinfo {author} {\bibfnamefont {S.}~\bibnamefont {Takahashi}}, \bibinfo
  {author} {\bibfnamefont {J.}~\bibnamefont {Ohe}}, \bibinfo {author}
  {\bibfnamefont {K.}~\bibnamefont {Uchida}}, \bibinfo {author} {\bibfnamefont
  {M.}~\bibnamefont {Mizuguchi}}, \bibinfo {author} {\bibfnamefont
  {H.}~\bibnamefont {Umezawa}}, \bibinfo {author} {\bibfnamefont
  {H.}~\bibnamefont {Kawai}}, \bibinfo {author} {\bibfnamefont
  {K.}~\bibnamefont {Ando}}, \bibinfo {author} {\bibfnamefont {K.}~\bibnamefont
  {Takanashi}}, \bibinfo {author} {\bibfnamefont {S.}~\bibnamefont {Maekawa}},
  \ and\ \bibinfo {author} {\bibfnamefont {E.}~\bibnamefont {Saitoh}},\ }\href
  {\doibase 10.1038/nature08876} {\bibfield  {journal} {\bibinfo  {journal}
  {Nature (London)}\ }\textbf {\bibinfo {volume} {464}},\ \bibinfo {pages}
  {262} (\bibinfo {year} {2010})}\BibitemShut {NoStop}%
\bibitem [{\citenamefont {Yoshino}\ \emph {et~al.}(2011)\citenamefont
  {Yoshino}, \citenamefont {Ando}, \citenamefont {Harii}, \citenamefont
  {Nakayama}, \citenamefont {Kajiwara},\ and\ \citenamefont
  {Saitoh}}]{SCNiFeratio}%
  \BibitemOpen
  \bibfield  {author} {\bibinfo {author} {\bibfnamefont {T.}~\bibnamefont
  {Yoshino}}, \bibinfo {author} {\bibfnamefont {K.}~\bibnamefont {Ando}},
  \bibinfo {author} {\bibfnamefont {K.}~\bibnamefont {Harii}}, \bibinfo
  {author} {\bibfnamefont {H.}~\bibnamefont {Nakayama}}, \bibinfo {author}
  {\bibfnamefont {Y.}~\bibnamefont {Kajiwara}}, \ and\ \bibinfo {author}
  {\bibfnamefont {E.}~\bibnamefont {Saitoh}},\ }\href {\doibase
  10.1063/1.3571556} {\bibfield  {journal} {\bibinfo  {journal} {Applied
  Physics Letters}\ }\textbf {\bibinfo {volume} {98}},\ \bibinfo {eid} {132503}
  (\bibinfo {year} {2011})}\BibitemShut {NoStop}%
\bibitem [{\citenamefont {Ando}\ \emph {et~al.}(2010)\citenamefont {Ando},
  \citenamefont {Kajiwara}, \citenamefont {Sasage}, \citenamefont {Uchida},\
  and\ \citenamefont {Saitoh}}]{AndoIEEE2010}%
  \BibitemOpen
  \bibfield  {author} {\bibinfo {author} {\bibfnamefont {K.}~\bibnamefont
  {Ando}}, \bibinfo {author} {\bibfnamefont {Y.}~\bibnamefont {Kajiwara}},
  \bibinfo {author} {\bibfnamefont {K.}~\bibnamefont {Sasage}}, \bibinfo
  {author} {\bibfnamefont {K.}~\bibnamefont {Uchida}}, \ and\ \bibinfo {author}
  {\bibfnamefont {E.}~\bibnamefont {Saitoh}},\ }\href@noop {} {\bibfield
  {journal} {\bibinfo  {journal} {IEEE Transactions on Magnetics}\ }\textbf
  {\bibinfo {volume} {46}},\ \bibinfo {pages} {3694 } (\bibinfo {year}
  {2010})}\BibitemShut {NoStop}%
\bibitem [{\citenamefont {Nakayama}\ \emph {et~al.}(2011)\citenamefont
  {Nakayama}, \citenamefont {Ando}, \citenamefont {Harii}, \citenamefont
  {Fujikawa}, \citenamefont {Kajiwara}, \citenamefont {Yoshino},\ and\
  \citenamefont {Saitoh}}]{1742-6596-266-1-012100}%
  \BibitemOpen
  \bibfield  {author} {\bibinfo {author} {\bibfnamefont {H.}~\bibnamefont
  {Nakayama}}, \bibinfo {author} {\bibfnamefont {K.}~\bibnamefont {Ando}},
  \bibinfo {author} {\bibfnamefont {K.}~\bibnamefont {Harii}}, \bibinfo
  {author} {\bibfnamefont {Y.}~\bibnamefont {Fujikawa}}, \bibinfo {author}
  {\bibfnamefont {Y.}~\bibnamefont {Kajiwara}}, \bibinfo {author}
  {\bibfnamefont {T.}~\bibnamefont {Yoshino}}, \ and\ \bibinfo {author}
  {\bibfnamefont {E.}~\bibnamefont {Saitoh}},\ }\href
  {http://stacks.iop.org/1742-6596/266/i=1/a=012100} {\bibfield  {journal}
  {\bibinfo  {journal} {Journal of Physics: Conference Series}\ }\textbf
  {\bibinfo {volume} {266}},\ \bibinfo {pages} {012100} (\bibinfo {year}
  {2011})}\BibitemShut {NoStop}%
\bibitem [{\citenamefont {Ando}\ \emph {et~al.}(2011)\citenamefont {Ando},
  \citenamefont {Takahashi}, \citenamefont {Ieda}, \citenamefont {Kajiwara},
  \citenamefont {Nakayama}, \citenamefont {Yoshino}, \citenamefont {Harii},
  \citenamefont {Fujikawa}, \citenamefont {Matsuo}, \citenamefont {Maekawa},\
  and\ \citenamefont {Saitoh}}]{VariousSizeAndo2011}%
  \BibitemOpen
  \bibfield  {author} {\bibinfo {author} {\bibfnamefont {K.}~\bibnamefont
  {Ando}}, \bibinfo {author} {\bibfnamefont {S.}~\bibnamefont {Takahashi}},
  \bibinfo {author} {\bibfnamefont {J.}~\bibnamefont {Ieda}}, \bibinfo {author}
  {\bibfnamefont {Y.}~\bibnamefont {Kajiwara}}, \bibinfo {author}
  {\bibfnamefont {H.}~\bibnamefont {Nakayama}}, \bibinfo {author}
  {\bibfnamefont {T.}~\bibnamefont {Yoshino}}, \bibinfo {author} {\bibfnamefont
  {K.}~\bibnamefont {Harii}}, \bibinfo {author} {\bibfnamefont
  {Y.}~\bibnamefont {Fujikawa}}, \bibinfo {author} {\bibfnamefont
  {M.}~\bibnamefont {Matsuo}}, \bibinfo {author} {\bibfnamefont
  {S.}~\bibnamefont {Maekawa}}, \ and\ \bibinfo {author} {\bibfnamefont
  {E.}~\bibnamefont {Saitoh}},\ }\href {\doibase 10.1063/1.3587173} {\bibfield
  {journal} {\bibinfo  {journal} {Journal of Applied Physics}\ }\textbf
  {\bibinfo {volume} {109}},\ \bibinfo {eid} {103913} (\bibinfo {year}
  {2011})}\BibitemShut {NoStop}%
\bibitem [{\citenamefont {Azevedo}\ \emph {et~al.}(2011)\citenamefont
  {Azevedo}, \citenamefont {Vilela-Le\~ao}, \citenamefont
  {Rodr\'iguez-Su\'arez}, \citenamefont {Lacerda~Santos},\ and\ \citenamefont
  {Rezende}}]{AzevedoPRB2011}%
  \BibitemOpen
  \bibfield  {author} {\bibinfo {author} {\bibfnamefont {A.}~\bibnamefont
  {Azevedo}}, \bibinfo {author} {\bibfnamefont {L.~H.}\ \bibnamefont
  {Vilela-Le\~ao}}, \bibinfo {author} {\bibfnamefont {R.~L.}\ \bibnamefont
  {Rodr\'iguez-Su\'arez}}, \bibinfo {author} {\bibfnamefont {A.~F.}\
  \bibnamefont {Lacerda~Santos}}, \ and\ \bibinfo {author} {\bibfnamefont
  {S.~M.}\ \bibnamefont {Rezende}},\ }\href {\doibase
  10.1103/PhysRevB.83.144402} {\bibfield  {journal} {\bibinfo  {journal} {Phys.
  Rev. B}\ }\textbf {\bibinfo {volume} {83}},\ \bibinfo {pages} {144402}
  (\bibinfo {year} {2011})}\BibitemShut {NoStop}%
\bibitem [{\citenamefont {Nakayama}\ \emph {et~al.}(2012)\citenamefont
  {Nakayama}, \citenamefont {Ando}, \citenamefont {Harii}, \citenamefont
  {Yoshino}, \citenamefont {Takahashi}, \citenamefont {Kajiwara}, \citenamefont
  {Uchida}, \citenamefont {Fujikawa},\ and\ \citenamefont
  {Saitoh}}]{NiFePtdep}%
  \BibitemOpen
  \bibfield  {author} {\bibinfo {author} {\bibfnamefont {H.}~\bibnamefont
  {Nakayama}}, \bibinfo {author} {\bibfnamefont {K.}~\bibnamefont {Ando}},
  \bibinfo {author} {\bibfnamefont {K.}~\bibnamefont {Harii}}, \bibinfo
  {author} {\bibfnamefont {T.}~\bibnamefont {Yoshino}}, \bibinfo {author}
  {\bibfnamefont {R.}~\bibnamefont {Takahashi}}, \bibinfo {author}
  {\bibfnamefont {Y.}~\bibnamefont {Kajiwara}}, \bibinfo {author}
  {\bibfnamefont {K.}~\bibnamefont {Uchida}}, \bibinfo {author} {\bibfnamefont
  {Y.}~\bibnamefont {Fujikawa}}, \ and\ \bibinfo {author} {\bibfnamefont
  {E.}~\bibnamefont {Saitoh}},\ }\href {\doibase 10.1103/PhysRevB.85.144408}
  {\bibfield  {journal} {\bibinfo  {journal} {Phys. Rev. B}\ }\textbf {\bibinfo
  {volume} {85}},\ \bibinfo {pages} {144408} (\bibinfo {year}
  {2012})}\BibitemShut {NoStop}%
\bibitem [{\citenamefont {Feng}\ \emph {et~al.}(2012)\citenamefont {Feng},
  \citenamefont {Hu}, \citenamefont {Sun}, \citenamefont {You}, \citenamefont
  {Wu}, \citenamefont {Du}, \citenamefont {Zhang}, \citenamefont {Hu},
  \citenamefont {Yang}, \citenamefont {Tang}, \citenamefont {Zhang},\ and\
  \citenamefont {Ding}}]{Feng2012}%
  \BibitemOpen
  \bibfield  {author} {\bibinfo {author} {\bibfnamefont {Z.}~\bibnamefont
  {Feng}}, \bibinfo {author} {\bibfnamefont {J.}~\bibnamefont {Hu}}, \bibinfo
  {author} {\bibfnamefont {L.}~\bibnamefont {Sun}}, \bibinfo {author}
  {\bibfnamefont {B.}~\bibnamefont {You}}, \bibinfo {author} {\bibfnamefont
  {D.}~\bibnamefont {Wu}}, \bibinfo {author} {\bibfnamefont {J.}~\bibnamefont
  {Du}}, \bibinfo {author} {\bibfnamefont {W.}~\bibnamefont {Zhang}}, \bibinfo
  {author} {\bibfnamefont {A.}~\bibnamefont {Hu}}, \bibinfo {author}
  {\bibfnamefont {Y.}~\bibnamefont {Yang}}, \bibinfo {author} {\bibfnamefont
  {D.~M.}\ \bibnamefont {Tang}}, \bibinfo {author} {\bibfnamefont {B.~S.}\
  \bibnamefont {Zhang}}, \ and\ \bibinfo {author} {\bibfnamefont {H.~F.}\
  \bibnamefont {Ding}},\ }\href {\doibase 10.1103/PhysRevB.85.214423}
  {\bibfield  {journal} {\bibinfo  {journal} {Phys. Rev. B}\ }\textbf {\bibinfo
  {volume} {85}},\ \bibinfo {pages} {214423} (\bibinfo {year}
  {2012})}\BibitemShut {NoStop}%
\bibitem [{\citenamefont {Kurebayashi}\ \emph {et~al.}(2011)\citenamefont
  {Kurebayashi}, \citenamefont {Dzyapko}, \citenamefont {Demidov},
  \citenamefont {Fang}, \citenamefont {Ferguson},\ and\ \citenamefont
  {Demokritov}}]{Kurebayashi2011nmat}%
  \BibitemOpen
  \bibfield  {author} {\bibinfo {author} {\bibfnamefont {H.}~\bibnamefont
  {Kurebayashi}}, \bibinfo {author} {\bibfnamefont {O.}~\bibnamefont
  {Dzyapko}}, \bibinfo {author} {\bibfnamefont {V.~E.}\ \bibnamefont
  {Demidov}}, \bibinfo {author} {\bibfnamefont {D.}~\bibnamefont {Fang}},
  \bibinfo {author} {\bibfnamefont {A.~J.}\ \bibnamefont {Ferguson}}, \ and\
  \bibinfo {author} {\bibfnamefont {S.~O.}\ \bibnamefont {Demokritov}},\
  }\href@noop {} {\bibfield  {journal} {\bibinfo  {journal} {Nat. Mater.}\
  }\textbf {\bibinfo {volume} {10}},\ \bibinfo {pages} {660} (\bibinfo {year}
  {2011})}\BibitemShut {NoStop}%
\bibitem [{\citenamefont {Tserkovnyak}, \citenamefont {Brataas},\ and\
  \citenamefont {Bauer}(2002)}]{PhysRevLett.88.117601}%
  \BibitemOpen
  \bibfield  {author} {\bibinfo {author} {\bibfnamefont {Y.}~\bibnamefont
  {Tserkovnyak}}, \bibinfo {author} {\bibfnamefont {A.}~\bibnamefont
  {Brataas}}, \ and\ \bibinfo {author} {\bibfnamefont {G.~E.~W.}\ \bibnamefont
  {Bauer}},\ }\href {\doibase 10.1103/PhysRevLett.88.117601} {\bibfield
  {journal} {\bibinfo  {journal} {Phys. Rev. Lett.}\ }\textbf {\bibinfo
  {volume} {88}},\ \bibinfo {pages} {117601} (\bibinfo {year}
  {2002})}\BibitemShut {NoStop}%
\bibitem [{\citenamefont {Tserkovnyak}, \citenamefont {Brataas},\ and\
  \citenamefont {Bauer}(2003)}]{PhysRevB.67.140404}%
  \BibitemOpen
  \bibfield  {author} {\bibinfo {author} {\bibfnamefont {Y.}~\bibnamefont
  {Tserkovnyak}}, \bibinfo {author} {\bibfnamefont {A.}~\bibnamefont
  {Brataas}}, \ and\ \bibinfo {author} {\bibfnamefont {G.~E.~W.}\ \bibnamefont
  {Bauer}},\ }\href {\doibase 10.1103/PhysRevB.67.140404} {\bibfield  {journal}
  {\bibinfo  {journal} {Phys. Rev. B}\ }\textbf {\bibinfo {volume} {67}},\
  \bibinfo {pages} {140404} (\bibinfo {year} {2003})}\BibitemShut {NoStop}%
\bibitem [{\citenamefont {Sandweg}\ \emph {et~al.}(2011)\citenamefont
  {Sandweg}, \citenamefont {Kajiwara}, \citenamefont {Chumak}, \citenamefont
  {Serga}, \citenamefont {Vasyuchka}, \citenamefont {Jungfleisch},
  \citenamefont {Saitoh},\ and\ \citenamefont
  {Hillebrands}}]{HillebrandsSPmagnons}%
  \BibitemOpen
  \bibfield  {author} {\bibinfo {author} {\bibfnamefont {C.~W.}\ \bibnamefont
  {Sandweg}}, \bibinfo {author} {\bibfnamefont {Y.}~\bibnamefont {Kajiwara}},
  \bibinfo {author} {\bibfnamefont {A.~V.}\ \bibnamefont {Chumak}}, \bibinfo
  {author} {\bibfnamefont {A.~A.}\ \bibnamefont {Serga}}, \bibinfo {author}
  {\bibfnamefont {V.~I.}\ \bibnamefont {Vasyuchka}}, \bibinfo {author}
  {\bibfnamefont {M.~B.}\ \bibnamefont {Jungfleisch}}, \bibinfo {author}
  {\bibfnamefont {E.}~\bibnamefont {Saitoh}}, \ and\ \bibinfo {author}
  {\bibfnamefont {B.}~\bibnamefont {Hillebrands}},\ }\href {\doibase
  10.1103/PhysRevLett.106.216601} {\bibfield  {journal} {\bibinfo  {journal}
  {Phys. Rev. Lett.}\ }\textbf {\bibinfo {volume} {106}},\ \bibinfo {pages}
  {216601} (\bibinfo {year} {2011})}\BibitemShut {NoStop}%
\bibitem [{\citenamefont {Castel}\ \emph {et~al.}(2012)\citenamefont {Castel},
  \citenamefont {Youssef}, \citenamefont {Vlietstra},\ and\ \citenamefont {van
  Wees}}]{YIGPt01}%
  \BibitemOpen
  \bibfield  {author} {\bibinfo {author} {\bibfnamefont {V.}~\bibnamefont
  {Castel}}, \bibinfo {author} {\bibfnamefont {J.~B.}\ \bibnamefont {Youssef}},
  \bibinfo {author} {\bibfnamefont {N.}~\bibnamefont {Vlietstra}}, \ and\
  \bibinfo {author} {\bibfnamefont {B.~J.}\ \bibnamefont {van Wees}},\
  }\href@noop {} {\enquote {\bibinfo {title} {{Frequency and power dependence
  of spin-current emission by spin pumping in a thin film YIG/Pt system}},}\ }
  (\bibinfo {year} {2012}),\ \Eprint {http://arxiv.org/abs/1206.6671v}
  {arXiv:1206.6671v} \BibitemShut {NoStop}%
\bibitem [{\citenamefont {Liu}, \citenamefont {Buhrman},\ and\ \citenamefont
  {Ralph}(2012)}]{RevPt}%
  \BibitemOpen
  \bibfield  {author} {\bibinfo {author} {\bibfnamefont {L.}~\bibnamefont
  {Liu}}, \bibinfo {author} {\bibfnamefont {R.~A.}\ \bibnamefont {Buhrman}}, \
  and\ \bibinfo {author} {\bibfnamefont {D.~C.}\ \bibnamefont {Ralph}},\
  }\href@noop {} {\enquote {\bibinfo {title} {{Review and Analysis of
  Measurements of the Spin Hall Effect in Platinum}},}\ } (\bibinfo {year}
  {2012}),\ \Eprint {http://arxiv.org/abs/1111.3702v3} {arXiv:1111.3702v3}
  \BibitemShut {NoStop}%
\bibitem [{\citenamefont {Mosendz}\ \emph {et~al.}(2010)\citenamefont
  {Mosendz}, \citenamefont {Pearson}, \citenamefont {Fradin}, \citenamefont
  {Bauer}, \citenamefont {Bader},\ and\ \citenamefont
  {Hoffmann}}]{PhysRevLett.104.046601}%
  \BibitemOpen
  \bibfield  {author} {\bibinfo {author} {\bibfnamefont {O.}~\bibnamefont
  {Mosendz}}, \bibinfo {author} {\bibfnamefont {J.~E.}\ \bibnamefont
  {Pearson}}, \bibinfo {author} {\bibfnamefont {F.~Y.}\ \bibnamefont {Fradin}},
  \bibinfo {author} {\bibfnamefont {G.~E.~W.}\ \bibnamefont {Bauer}}, \bibinfo
  {author} {\bibfnamefont {S.~D.}\ \bibnamefont {Bader}}, \ and\ \bibinfo
  {author} {\bibfnamefont {A.}~\bibnamefont {Hoffmann}},\ }\href {\doibase
  10.1103/PhysRevLett.104.046601} {\bibfield  {journal} {\bibinfo  {journal}
  {Phys. Rev. Lett.}\ }\textbf {\bibinfo {volume} {104}},\ \bibinfo {pages}
  {046601} (\bibinfo {year} {2010})}\BibitemShut {NoStop}%
\bibitem [{\citenamefont {Jungfleisch}\ \emph {et~al.}(2011)\citenamefont
  {Jungfleisch}, \citenamefont {Chumak}, \citenamefont {Vasyuchka},
  \citenamefont {Serga}, \citenamefont {Obry}, \citenamefont {Schultheiss},
  \citenamefont {Beck}, \citenamefont {Karenowska}, \citenamefont {Saitoh},\
  and\ \citenamefont {Hillebrands}}]{2mag}%
  \BibitemOpen
  \bibfield  {author} {\bibinfo {author} {\bibfnamefont {M.~B.}\ \bibnamefont
  {Jungfleisch}}, \bibinfo {author} {\bibfnamefont {A.~V.}\ \bibnamefont
  {Chumak}}, \bibinfo {author} {\bibfnamefont {V.~I.}\ \bibnamefont
  {Vasyuchka}}, \bibinfo {author} {\bibfnamefont {A.~A.}\ \bibnamefont
  {Serga}}, \bibinfo {author} {\bibfnamefont {B.}~\bibnamefont {Obry}},
  \bibinfo {author} {\bibfnamefont {H.}~\bibnamefont {Schultheiss}}, \bibinfo
  {author} {\bibfnamefont {P.~A.}\ \bibnamefont {Beck}}, \bibinfo {author}
  {\bibfnamefont {A.~D.}\ \bibnamefont {Karenowska}}, \bibinfo {author}
  {\bibfnamefont {E.}~\bibnamefont {Saitoh}}, \ and\ \bibinfo {author}
  {\bibfnamefont {B.}~\bibnamefont {Hillebrands}},\ }\href {\doibase
  10.1063/1.3658398} {\bibfield  {journal} {\bibinfo  {journal} {Applied
  Physics Letters}\ }\textbf {\bibinfo {volume} {99}},\ \bibinfo {eid} {182512}
  (\bibinfo {year} {2011})}\BibitemShut {NoStop}%
\bibitem [{\citenamefont {Jia}\ \emph {et~al.}(2011)\citenamefont {Jia},
  \citenamefont {Liu}, \citenamefont {Xia},\ and\ \citenamefont
  {Bauer}}]{GerritMixCon}%
  \BibitemOpen
  \bibfield  {author} {\bibinfo {author} {\bibfnamefont {X.}~\bibnamefont
  {Jia}}, \bibinfo {author} {\bibfnamefont {K.}~\bibnamefont {Liu}}, \bibinfo
  {author} {\bibfnamefont {K.}~\bibnamefont {Xia}}, \ and\ \bibinfo {author}
  {\bibfnamefont {G.~E.~W.}\ \bibnamefont {Bauer}},\ }\href
  {http://stacks.iop.org/0295-5075/96/i=1/a=17005} {\bibfield  {journal}
  {\bibinfo  {journal} {EPL (Europhysics Letters)}\ }\textbf {\bibinfo {volume}
  {96}},\ \bibinfo {pages} {17005} (\bibinfo {year} {2011})}\BibitemShut
  {NoStop}%
\end{thebibliography}

\providecommand{\noopsort}[1]{}\providecommand{\singleletter}[1]{#1}%

\end{document}